\documentclass[aps,prb,twocolumn,10pt]{revtex4}
\usepackage[dvips]{graphicx}
\usepackage{subfig}
\usepackage[usenames,dvipsnames]{pstricks}
\usepackage{epsfig}
\begin{document}
\title{\textsf{Exact eigenstate analysis of finite-frequency conductivity in graphene}}
\author{Rajyavardhan Ray}
\author{Avinash Singh}
\email{avinas@iitk.ac.in}
\affiliation{Department of Physics, Indian Institute of Technology Kanpur, India}

\begin{abstract}
We employ the exact eigenstate basis formalism to study electrical conductivity in graphene, in the presence of short-range diagonal disorder and inter-valley scattering. We find that for disorder strength, $W \ge$ 5, the density of states is flat. We, then, make connection, using  the MRG approach, with the work of Abrahams \textit{et al.} and find a very good agreement for disorder strength, $W$ = 5. For low disorder strength, $W$ = 2, we plot the energy-resolved current matrix elements squared for different locations of the Fermi energy from the band centre. We find that the states close to the band centre are more extended and falls of nearly as $1/E_l^{2}$ as we move away from the band centre. Further studies of current matrix elements versus disorder strength suggests a cross-over from weakly localized to a very weakly localized system. We calculate conductivity using Kubo Greenwood formula and show that, for low disorder strength, conductivity is in a good qualitative agreement with the experiments, even for the on-site disorder. The intensity plots of the eigenstates also reveal clear signatures of puddle formation for very small carrier concentration. We also make comparison with square lattice and find that graphene is more easily localized when subject to disorder.
\end{abstract}
\maketitle

\section{Introduction}
Graphene has been long studied as a theoretical toy model not only to understand it's appealing physical properties,\cite{Jakiw,Semenoff,Fradkin3,Haldane} but also as a basic building block of various Carbon allotropes like graphite,\cite{Wallace,Weiss} and more recently fullerenes and nanotubes.\cite{Castroneto2} 
While graphite is the three dimensional allotrope of Carbon and could be formed by the Bernal stacking of graphene sheets, fullerene and nanotubes are the zero and one dimensional allotropes, formed by introducing pentagonal impurities and rolling the graphene sheets, respectively. After its experimental isolation in 2004,\cite{Novoselov,Novoselov2, Zhang 2005} there has been a renewed interest in studying various properties of graphene sheet, both theoretically and experimentally, as well as due to potential technological applications.\cite{McD,Castroneto2,Katsnelson}

Graphene consists of a single sheet of Carbon atoms arranged on a honeycomb lattice. Basic properties of graphene are well described by a tight-binding model for the $\pi$-orbitals which are perpendicular to the graphene plane at each Carbon atom. The effective low-energy theory states that the charge carriers in graphene are massless Dirac fermions, characterized by a linear dispersion relation and a linear energy dependence of the density of states which vanishes at the Fermi level implying a semi-metallic behaviour for graphene.\cite{Wallace, Weiss, Geim} 
Graphene has attracted a lot of attention recently not only due to its potential technological applications but also for understanding of physics in 2D systems\cite{Castroneto2,Katsnelson}. Its low energy description mimics (2+1)-dimensional quantum electrodynamics and hence graphene could act as a testing ground for various relativistic phenomena.\cite{Katsnelson}

Early experiments on graphene have revealed that the conductivity at low temperatures is directly proportional to the carrier concentration (or gate voltage) except for very low carrier concentration. For zero gate voltage, the conductivity approaches a robust minimum universal value proportional to $e^{2}/h$.\cite{Novoselov,Zhang 2005} This could not be explained by the Born approximation which predicts a conductivity independent of carrier concentration.\cite{Shon} Other interesting properties include anomalous integral quantum Hall effect and suppression of weak localization.\cite{Castroneto2} Recent experiments, however, show that the dependence of conductivity on carrier concentration could vary from sub linear to superlinear for different carrier concentrations.\cite{Zhang 2005,Morozov 2008} It has been argued that presence of impurities in graphene is the main contributor towards its electronic properties.\cite{Wehling} The importance of disorder in graphene could most easily be emphasized by observing that the universal conductivity suggested by the theoretical studies on defectless graphene sheet is 2-20 times smaller than the observed conductivity close to the Dirac points.\cite{Rossi}

The Boltzmann conductivity for graphene is given by $\sigma_0 = (e^2/h)(2 E_F \tau_0 / \hbar)$. The observed conductivity rises linearly with carrier concentration in graphene and $N(E_{F}) \propto \sqrt n$ , where $N(E_{F})$ is the density of states at the Fermi energy and $n$ is the carrier density. This implies that the scattering rate, $\tau(E_{F}) \propto \sqrt n$. On the other hand, for weak local scatterers, Born approximation predicts $\tau ^{-1} \propto n_{\rm imp}N(E_{F}) \propto \sqrt n$ where $n_{\rm imp}$ is the impurity concentration.\cite{Shon} In view of this discrepancy, various investigations, both theoretical and numerical, have been carried out in order to understand the behavior of graphene under various types of disorder,\cite{Altland, Ostro, Amini, Xiong, Khveshchenko, Pereira, Zeigler, Castroneto2, Guinea, Adam, Katsnelson2, McD, Pereira MD, Zhang, Lewenkopf, Hwang, Pereira CI, Rossi, Zeigler2, Vyurkov} such as vacancies,\cite{Pereira,Pereira MD} charged carriers,\cite{McD,Hwang,Pereira CI,Vyurkov} on-site disorder,\cite{Pereira MD,Xiong} long range on-site disorder,\cite{Lewenkopf} off-diagonal disorder,\cite{Xiong,Amini} off-diagonal disorder with sign change probability in the hopping term.\cite{Xiong}

Vacancies have been proposed to induce localized states, extended over many lattice sites, which are sensitive to the electron-hole symmetry breaking\cite{Pereira} Detailed studies in the presence of both compensated and uncompensated defects reveal that they could modify the low energy spectrum in graphene drastically like there could be quasi-localized zero modes and introduction of gap in the DOS.\cite{Pereira MD} 

For charged scatterers, Nomura \textit{et al.} \cite{McD} have argued on the basis of Boltzmann transport theory that the linear dependence of conductivity on carrier concentration could be explained. They find that states close to the Dirac point are delocalized leading to $\sigma_{\rm min}$. Also, one could observe antilocalization if the inter-valley scattering is weak. On the other hand, if inter-vally scattering is large, all states could be localized due to accumulation of Berry phases. Conductivity in the presence of random charged impurity is also studied by Hwang \textit{et al.}\cite{Hwang} 
They find linear dependence of conductivity on carrier concentration for
high carrier density. However, for low carrier density, they argue, that
system develops some inhomogenities (random electron-hole puddles) which
implies that this domain is dominated by localization physics. They also
conclude that change of bias voltage may change the average distance
between graphene sheet and the impurity in the substrate which could
lead to sub- and super-linear conductivity dependence on carrier
concentraion. Pereira \textit{et al.} \cite{Pereira CI} have argued that
there could be a ``critical coupling'' distinguishing strong and weak
coupling regimes in the presence of unscreened Coulomb charges. They
also find bound states and strong renormalization of Van Hove
singularities in the DOS. Vyurkov \textit{et al.} \cite{Vyurkov}have
argued that the intrinsic conductivity of graphene (ambipolar system) is
dominated by strong electron-hole scattering. It has a universal value
independent of temperature. It is shown that conductivity could be
proportional to V or $V^2$ depending on the other scattering mechanisms
present like those on phonons by charged defects. In the unipolar
system, it is argued that electron-hole scattering is not important and
conductivity is proportional to $V^{1/2}$. Trushin \textit{et al.}
\cite{Trushin} have
reported an analytical calculation for Boltzmann conductivity
with  screened Coulomb scatterers and both electron-hole coherent and
incoherent solutions. They find that the experimentally observed
dependence of conductivity on $n$ could be explained by the
electron-hole coherent solution.

For diagonal disorder, it is found that beyond a certain threshold disorder strength, bound states appear beyond the band continuum and resonant states could appear at low energies. In the infinite disorder strength limit, results match with that of vacancies. In the off-diagonal disorder case, strong low-energy resonances appear. There are, however, no bound states.\cite{Pereira MD} Localization studies on graphene with on-site disorder carried out by Xiong \textit{et al.}\cite{Xiong} reveal that all states are localized in the case of random diagonal disorder which is consistent with Anderson localization. Lherbier \textit{et al.}\cite{Lherbier} have also studied the energy dependent  elastic mean free path, charge mobilities and semi-classical conductivity in the presence of Anderson-type disorder, using real space order N Kubo formalism, for both two-dimensional graphene and graphene nano-ribbons (GNRs). It was found that the systems undergo a conventioanl Anderson transition in the zero temperature limit. 

Lewenkopf \textit{et al.} \cite{Lewenkopf} have studied long range diagonal disorder (Gaussian scatteres) at finite concentration. They find that conductivity increases as disorder strength is decreased. It is shown that conductivity depends only on disorder strength and ratio of the system size to disorder correlation length. This dependence could vary between sublinear to superlinear depending on disorder strength. For fixed disorder strength, conductivity increases with doping concentration. In the presence of strong long-range impurities, Zhang \textit{et al.}\cite{Zhang} have shown that states close to the Dirac points are localized for sufficiently strong disorder strength and Kosterlitz-Thouless transition between localized and delocalized states is proposed which is seen in terms of the ``current flow vector``.

In the case of random off-diagonal disorder (hopping disorder), localization studies by Xiong \textit{et al.}\cite{Xiong} reveal that states close to the Dirac point are delocalized due to chiral symmetry. They find that the off-diagonal disorder leads to a shape-dependent conductivity depending on the length to width ratio. However, if a sign change probability is introduced, they find that the conductivity becomes shape independent. Amini \textit{et al.}\cite{Amini} have shown that for  disorder strength less than the hopping strength i.e. $W < t$, there is no localization. Disorder $W$ slows down the Dirac quasi particles but preserves their nature. For $W \sim t$, localization sets in for states close to the Fermi energy, gap at energy close to the Dirac point and for $W > t$ existence of mobility edge is proposed which starts at Fermi energy and moves towards the edges. States close to the Fermi energy are extended. They also propose the existence of disorder induced gap defined as the distance between the upper and lower mobility edges around the Fermi point.

It should be noted that in all these works, the inter-valley scattering
is assumed to be very small and hence not contributing towards
conductivity. However, Klos \textit{et al.}\cite{Klos} have done a
comparative study of conductivity using the tight-binding(TB) Landauer
approach and on the basis of the Boltzmann theory and find a discrepancy
between that results obtained by TB calculation and Boltzmann approach.

Despite all the efforts, the issue of localization in graphene is currently highly debated from a theoretical standpoint. The obsered minimal conductivity $\sigma_{\rm min} \sim 4e^{2}/h$ over a range of mobilities remains to be fully understod. In view of a recent experiment by Ponomarenko \textit{et al.}\cite{Ponomarenko} and Katoch \textit{et al.}\cite{Katoch} , which explores the dominant scatterers in graphene,and Horng\textit{et al.}\cite{Horng}, which measures the high-frequency conductivity in graphene, unlike believed so far, the primary reason for the linear rise of conductivity with carrier concentration is also debatable.

In the present work, we will investigate the finite-frequency electrical
conductivity and localization properties of graphene in the presence of
diagonal (on-site) disorder for various disorder strengths. Our
exact-eigenstates approach implicitly takes into account the
inter-valley as well as the intra-valley scatterings. This paper is
organized into eight sections. 

In section II and III, we introduce the two sub-lattice basis and evaluate the exact single-particle Green's function within the t-matrix approach for a single impurity on either sublattice. 
In section IV, we consider disordered graphene. Disorder is introduced via random fluctuation of the on-site energies of the $\pi$-orbitals.

In section V, we use the Kubo-Greenwood formula to calculate the frequency-dependent conductivity for different disorder strengths and for different system sizes. The system-size dependence is employed to perform a renormalization group analysis, and for moderate disorder strength contact is made with the weak localization result of Abraham's \textit{et al.}. Later on, focus on the weak disorder strength. We study the frequency dependence of the averaged current matrix element squared for different locations in the band (Fermi energies) and hence calculate conductivity and mobility for low-disordered graphene samples. We also study the dependence of energy resolved current matrix elements squared on system size and disorder strength. 

In section VI, we study the average current matrix elements squared over a range of disorder strength and for different system sizes in order to gain a better insight into disorder induced localization in graphene. 

In section VII, we make a comparision between graphene and square lattice. We compare the normalized conductivity between the two. We also show the intensity plots for both the lattices for different values of disorder strength.

Finally, our conclusions are presented in section VIII.

\section{Formalism}
As already mentioned above, graphene is a single layer of Carbon atoms on a honeycomb lattice. A honeycomb lattice, which is not a Bravais lattice, could be viewed as a composed of two kinds of sub-lattices. Nearest neighbor hopping takes electron from one sub-lattice to another. We start with the tight-binding(TB) Hamiltonian:
\begin{equation}
 H = - t \sum_{<ij>} [a^{\dagger}_{j} a_{i} + c.c.]
\label{tbham}
\end{equation}
where the sum is over the nearest neightbours. This could be written in the two sub-lattice basis as:
\begin{equation}
\left[
\begin{array}{cc}
	0 & \delta(\mathbf k) \\
	\delta^{*}({\mathbf k}) & 0\\
\end{array}\right]
\label{ham1}
\end{equation}
where, the hopping term in k-space is:
\begin{equation}
\delta({\mathbf k}) = -t[e^{ik_x a} + 2e^{(-ik_{x}a / 2)}cos(\frac{\sqrt{3} k_y a}{2})]
\end{equation}
It should be noted that this Hamiltonian is ($2 \times 2$) and mixes the Dirac points. The energy eigenvalues are given by 
\begin{eqnarray}   
E_{\mathbf k}\!\!\!  &=&\!\!\! \pm \sqrt{\delta({\mathbf k})\delta^{*}(\mathbf k)} \nonumber\\
		   &=&\!\!\!  \pm t\! \sqrt{1\! +\! 4cos^2(\!\!\frac{\sqrt{3}k_y
               a}{2}\!\!)\! +\! 4cos(\!\!\frac{\sqrt{3}k_x
               a}{2}\!\!)cos(\!\!\frac{\sqrt{3}k_y a}{2}\!\!)}
\end{eqnarray}
In this basis, the low energy spectrum provides six Dirac points in the extended Brilluoin zone. These points are located at $ \vec k = (\pm 2\pi/3, \pm 2\pi/{3 \sqrt{3}})$ and $(0,\pm 4\pi/3\sqrt3)$. Expanding around the Dirac points, we find the linear dispersion for carriers in graphene.

\begin{figure}
\begin{center}
\subfloat[\label{fig:dos}]{\includegraphics[angle=-90,scale=0.32]{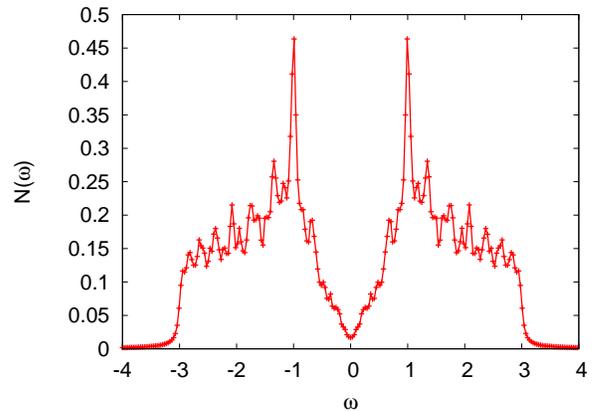}} \\
\vspace{-0.4cm}
\subfloat[\label{fig:gii}]{\includegraphics[angle=-90,scale=0.31]{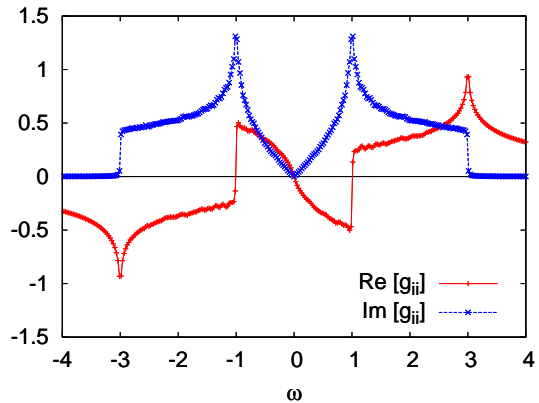}}
\end{center}
\vspace{-0.5cm}
\caption{(color online) Density of States in graphene: (a) Density of States and (b) Real and Imaginary parts of the Green's function.}
\label{fig:dos,gii}
\end{figure}

The free particle Green's function is given by the expression:
\begin{equation}
{\fontsize{8.7}{10}
G^{0}(\mathbf k,\omega)\! =\! \langle k |\frac{1}{\omega - H (\mathbf
k)}|k\rangle\!
    = \!\left[
\begin{array}{cc}
	\omega & \delta(\mathbf k)  \\
	\delta^{*}(\mathbf k) & \omega \\
\end{array}\right] \frac{1}{\omega^2\! -\! E_{\mathbf k}^2}
}
\end{equation}
where $E_{\mathbf k}$ is the energy eigenvalue derived earlier. The density of single particle states would then be given by
\begin{equation}
N(E)=\left\{
\begin{array}{rl}
	\frac{1}{\pi}{\sum_{\mathbf k}} Im [G^0(\mathbf k,\omega)] & \mbox{if }E > E_F  \\
	-\frac{1}{\pi}{\sum_{\mathbf k}} Im[G^0(\mathbf k,\omega)] & \mbox{if }E < E_F
\end{array}\right.
\end{equation}
Fig.(\ref{fig:dos,gii}) shows the density of states (DOS) for pure graphene (Fig.(\ref{fig:dos}))and also the real and imaginary part of the Green's function (Fig.(\ref{fig:gii})).

\section{T-MATIRX ANALYSIS}
\begin{figure}
\begin{center}
\includegraphics[scale=0.31]{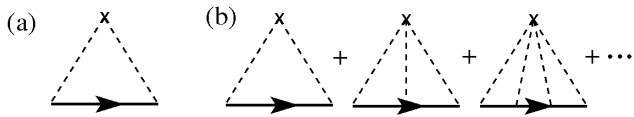}
\end{center}
\caption{Feynman diagrams for the  (a) Born and (b) $T$-matrix approximations for the self energy. The ``x'', dotted line and line with arrow signify the impurity, impurity potential and the Green's function respectively.}
\label{fig:t-matrix} 
\end{figure}

In the presence of impurities, the free particle Green's function gets modified. We shall, at this stage, look at the effect of a single impurity on the Green's function. We shall employ the T-matrix approach which gives exact result in the case of single impurity. As from the scattering theory, the modified Green's function based on the diagrams in Fig.(\ref{fig:t-matrix}) is be given by:
\begin{displaymath}
G_{ij}(\omega) = G^0_{ij}(\omega) + G^0_{iI}(w)T(\omega)G^0_{Ij}(\omega)
\end{displaymath}
where the T-matrix is given as $T(\omega)=V/[1-VG^0_{II}(\omega)]$, V is the strength of the impurity. The index $I(=A,B)$ and $(i,j)$ denotes the impurity site and the lattice sites respectively. 

Information regarding additional poles is contained in the T-matrix. This implies that there are possibilities of existence of new states if there is an intersection of $Re[G^0_{II}(\omega)]$ and 1/V. The Imaginary part is small but finite. If the strength of impurity is small or even moderate, there shall be no intersection with the $1/V$ line. It is only when the strength of the impurity is large, some intersection is expected and this might lead to a new state. Therefore, for some value of $\omega = \omega^{*}$, a resonance state occurs at the impurity site. This new state would be localized at the impurity site. The full Gren's function is given by:
\begin{equation}
G(\mathbf k,\omega) = \frac{[G^{0}(\mathbf k,\omega)]}{I - V[G^{0}_{II}(\omega)]}
\end{equation}

Around either $K$ or $K^{'}$ points (Dirac points), the $G^0_{II}$ has the following form.
\begin{eqnarray}
G^{0}_{II}(\omega) &=& \sum_{\mathbf k} \frac{\omega}{\omega^2 - E_{\mathbf k}^{2}} \nonumber\\ 
&=& - \frac{\omega \pi}{v_{\rm F}^2}[\ln(\frac{v_{\rm F}^{2}k^{2}_{c}-\omega^2}{\omega^2}) + i\pi]
\end{eqnarray}
where $E_{\mathbf k}$ is the energy, $k_c$ is the momentum cutoff, $v_{\rm F}$ is the Fermi velocity and $I$ is the impurity site. One should note that the free Green's function contributes equally for an impurity sitting on either sub-lattice A or B. Also, as, $\omega \rightarrow 0$, the $\omega$ term in front provides a cut-off to the log divergence. Imaginary part gives the density of states as already mentioned above. Real part of the above expression is small.

If we introduce a magnetic impurity at the some site $I$, there could again be appearance of new states for sufficiently large impurity strength. However, the intersection could now be either above or below the $\omega$ axis. However, since the nature of intersection is same as for the single non-magnetic impurity, the physics is expected to remain same.

\section{Disordered Graphene}
We will now include diagonal (on-site) disorder and consider the following tight-binding Hamiltonian:
\begin{equation}
H = \sum_{i,\sigma} \epsilon_{i} a^{\dagger}_{i \sigma} a_{i \sigma} - t \sum_{<ij> \sigma} [a^{\dagger}_{j \sigma} a_{i \sigma} + a^{\dagger}_{i \sigma} a_{j \sigma}]
\label{ham2}
\end{equation}
on a honeycomb lattice. Here the random on-site energies $\epsilon_i$ are chosen from a uniform distribution on $-W/2 \leq \epsilon_{i} \leq W/2$. The second term is the hopping term with summation over nearest-neighbour pairs of sites. From the matrix realization of the above Hamiltonian on a finite lattice with periodic boundary condition in both directions, we numerically obtain the eigenfunctions $\phi^{l}_{i}$ and the eigenvalues $E_l$ of the Hamiltonian in Eq.(\ref{ham2}) by exact diagonalization of the Hamiltonian matrix. One should note that the spin index simply runs through the calculation and shall be considered while making comparisions with the experiments. In the following, we have set the hopping term $t=1$ as the unit of energy.
\begin{figure}
\begin{center}
\includegraphics[angle=-90,scale=0.31]{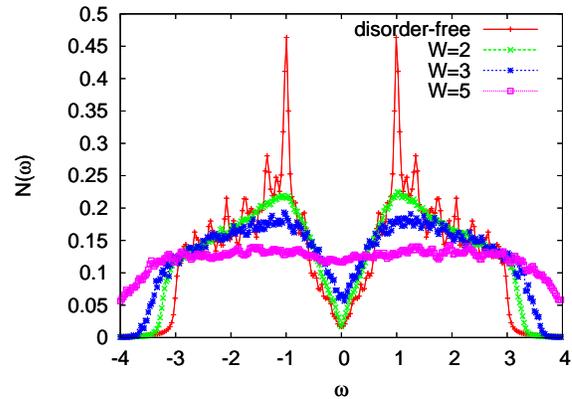}
\end{center}
\caption{(color online) Density of states for a graphene sheet with random on-site impurity potential. For, W=2, it is still linear. However, for W=5, it becomes nearly flat. Energies are measured in the units of $t$.}
\label{fig:dos_dis} 
\end{figure}

From the exact eigenvalues we obtain the local density of states (LDOS) using:
\begin{equation}
\frac{1}{N}\sum_{\mathbf k} \delta(E - E_{\mathbf k}) = \frac{1}{\pi N} \sum_{\mathbf k} \frac{\eta}{(E - E_{\mathbf k})^2 + \eta ^2} \nonumber
\end{equation}
where $N$ is the total number of lattice points in the system.

Fig.(\ref{fig:dos_dis}) shows the LDOS for various disorder strengths. Clearly, the van Hove singularities in the pure graphene DOS are softened with increasing disorder strength. For very small disorder strength, the DOS is still linear. However, for $W \ge 5$, the density of states becomes independent of $\omega$. This could be understood on the basis of intersection of $Re[G^0_{II}(w)]$ and 1/V, as argued in the previous section. 

\section{Finite-frequency conductivity }
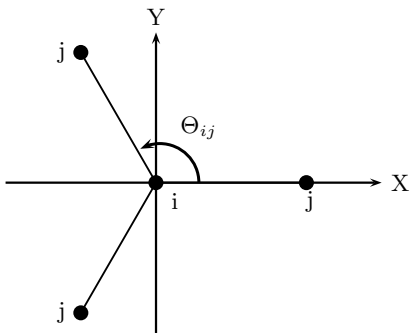
\begin{figure}
\begin{center}
\begin{pspicture}(2.3,-1.00)(4.3,3.05)
\psline[linewidth=0.03cm]{->}(1.00,1.00)(6.00,1.00)
\psline[linewidth=0.03cm]{->}(3.00,-1.00)(3.00,3.00)
\psline[linewidth=0.025cm](5.00,1.00)(3.00,1.00)
\psline[linewidth=0.025cm](3.00,1.00)(2.00,2.732)
\psline[linewidth=0.025cm](3.00,1.00)(2.00,-0.732)
\psdots[dotsize=0.2](5.00,1.00)
\psdots[dotsize=0.2](3.00,1.00)
\psdots[dotsize=0.2](2.00,2.732)
\psdots[dotsize=0.2](2.00,-0.732)
\psarc[linewidth=0.04]{->}(3.05,1.0){0.52}{0.0}{120.0}
\rput(3.57,1.73){$\Theta_{ij}$}
\rput(3.25,0.75){i}
\rput(5.05,0.75){j}
\rput(1.75,2.732){j}
\rput(1.75,-0.732){j}
\rput(6.25,1.00){X}
\rput(3.00,3.20){Y}
\end{pspicture}
\end{center}
\caption{Nearest neighbors to a Carbon atom on one of the sublattice on a honeycomb lattice. The dots represent C atoms. $\Theta_{ij}$ is the angle between the X-axis and the other nearest neighbors.}
\label{fig:theta}
\end{figure}

We are interested in calculating finite-frequency conductivity for graphene. We use the Kubo-Greenwood formula\cite{Kubo, Greenwood} for conductivity which uses the eigenvalues $E_l$ and the eigenfunctions $\phi^{l}_{i}$ calculated earlier. The Kubo-Greenwood formula is given by:
\begin{widetext}
\begin{equation}
\sigma_{xx}(\omega) =  \frac{\pi}{\omega N_u A}\int_{E_{F}-\hbar\omega}^{E_{F}}  dE \sum_{l,m} \vert M^{x}_{lm} (\omega) \vert ^{2} \delta(E - E_{m}) \delta(E + \hbar \omega - E_{m})
\label{KG}
\end{equation}
where, the current matrix element $M^{x}_{lm}(\omega)$ is given by:
\begin{eqnarray}
M^{\mu}_{lm}(\omega)&=&\langle l \vert J_{\mu} \vert m \rangle \nonumber\\
		    &=&\frac{ieat}{\hbar}\sum_{<ij>} (\phi ^{i} _{l} \phi
                ^{j} _{m} - \phi ^{j} _{l} \phi ^{i} _{m})\times \{
   		\begin{array}{rl}
		cos \Theta_{ij}&\mbox{for }\mu = x\\
		sin \Theta_{ij}&\mbox{for }\mu = y
		\end{array}
\end{eqnarray}
\end{widetext}
where $|l\rangle$ and $|m\rangle$ are the eigenstates with $E_{m} < E < E_{l}$ and $\Theta_{ij}$ is the angle between the $x$-axis and the two other nearest neighbors at each site as shown in Fig.(\ref{fig:theta}). Summation ,$<ij>$ is over nearest neighbors. Here $N_u$ and $A = 3 \sqrt{3}a^{2}/2$ are the number of unit cells and area of each unit cell, respectively.

The summation over $l,m$ in Eq.(\ref{KG}) is carried out by considering each pair of states $|l\rangle$ and $|m\rangle$ such that the energy difference, $\hbar \omega = E_l-E_m$, remains fixed. 

We shall first consider the case for $W$ = 5.

\subsection{W=5 }
When the density of states is independent of $\omega$ (refer Fig. \ref{fig:dos_dis}, W=5), the energy levels are nearly equally spaced. For a given random distribution of on-site impurity potentials, the contribution to the matrix elements from each energy pair of eigenvalues with fixed energy difference, $\hbar \omega = E_l-E_m$ could simply carried out by keeping $l-m$ fixed. We average the total contribution by the number of such paired states considered and the number of random configurations,typically 5000 samples of the matrix elements in all.

The expression for the finite-frequency conductivity then reduces to:
\begin{equation}
\sigma_{n}(\omega) = 8\pi \hbar N M_{n}^{2}(\omega)N(0)^{2}/3\sqrt{3}t^{2}a^{2}
\label{sigma}
\end{equation}
where $M_{n}^{2}(\omega)$ is the averaged matrix element squared and $N(0)$ is the averaged density of states per site at the Dirac points.

\begin{figure}
\begin{center}
\includegraphics[scale=0.40]{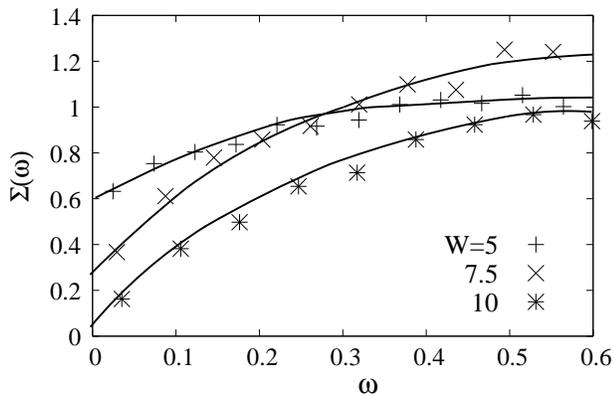}
\end{center}
\caption{(color online) The normalized conductivity, $\Sigma(\omega)$ versus $\omega$ for various impurity strengths for a (20,24) system. In all the cases, the DOS is flat. The curves are drawn as a guide to the eye. Some data points are omitted for clarity.}
\label{fig:normcond1} 
\end{figure}

\begin{figure}
\begin{center}
\includegraphics[angle=0,scale=0.40]{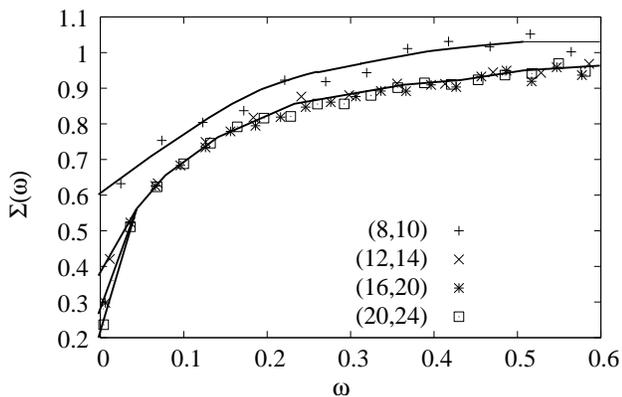}
\end{center}
\caption{(color online) The normalized conductivity, $\Sigma(\omega)$versus$\omega$ for various system sizes. In all the cases, the DOS is flat (W=5). The curves are drawn as a guide to the eye. Some data points are omitted for clarity.}
\label{fig:normcond2} 
\end{figure}

We now define the normalized conductivity 
\begin{equation}
\Sigma_{n}(\omega)=3\sqrt{3}\hbar W^{2} \sigma_{n}(\omega)/8\pi e^{2}
\end{equation}
which is convenient for plotting the data.  Fig. (\ref{fig:normcond1}) is the data for a (20,24) lattice for various values of $W$. At high frequency the normalized conductivity is approximately one. For higher $W$, the conductivity falls off at low frequency. This reduction of the low-frequency conductivity can be ascribed to the effects of localization. We show in Fig.(\ref{fig:normcond2}) the normalized conductivity versus frequency for different system sizes at fixed $W$($W=5$). We observe that the conductivity decreses with increasing system size which exhibits the renormalization group at work.

We also note that, despite the fact that the lattice is not symmetric about $x$ and $y$ axes, the conductivity behaviour turns out to be similar in both directions (not shown). This is due to the fact that the spectrum of graphene is conical. 

\subsubsection{Macroscopic Renormalization Group}
We now turn to the macroscopic renormalization-group (MRG) calculation. We start with a certain set of Hamiltonian parameters ($t$ and $W$) and calculate the appropriate macroscopic physical quantites as a function of lattice parameter. The hamiltonian parameters are then renormalized to preserve the physical quantites as the lattice parameter is varied. We consider specifically two systems with different lattice spacing ($a$ and $a^{'}$) but same physical size. For these two systems to represent the same physical problem with different microscopic length scales, we demand that the Hamiltonian parameters be so related that the physical properties are preserved. The appropriate physical quantites to be preserved for the localization problem are the one-electron DOS at the Fermi energy, $N(0)$ and the low-frequency electrical conductivity. \cite{AS}. We define the dimensionless conductance which depends only on $W$.
\begin{equation}
g_{n}(\omega) = \hbar \sigma_{n}(\omega)/e^{2}
\label{conductance}
\end{equation}

Then, the one-parameter RG recursion relation is:
\begin{equation}
g_{n}(\omega,w^{'})=g_{n}(\omega,w)
\label{mrg1}
\end{equation}
The other MRG relation:
\begin{equation}
N_{n}(t^{'},W^{'})=N_{n}(t,W)
\label{mrg2}
\end{equation}
fixes the absolute magnitude of $t$ to preserve the DOS. For weak disorder, the weak localization scaling theory result in 2D is\cite{Abrahams}
\begin{equation}
\partial \ln g / \partial \ln (L) = \beta(g) = -1/2\pi^{2}g
\end{equation}
where, L is the length scale. For $W=5$, we compare the (20,24) lattice with (12,14) lattice and find that $\Delta g/\Delta \ln (\sqrt{N}) = -0.052$ which matches well with the above value: $-1/2\pi^{2} = -0.051$. Thus, for $W=5$, the scaling theory of weak localization is completely obeyed.

\subsection{W=2}
We now turn our attention to the case when the density of states is not flat (e.g. for W=2, refer Fig. \ref{fig:dos_dis}). The earlier method of obtaining the average current matrix elements squared does not work here since the LDOS is linear. We, therefore, average the current matrix elements squared, $|M_{lm}|^{2}$ over all $l$ and $m$ states such that the energy difference $\hbar \omega=E_l - E_m$ lies in a given range (binning). We have checked that for $W$=5 the $<|M_{lm}|^{2}>$ obtained by this method matches with the earlier one.

\begin{figure}
\begin{center}
\includegraphics[angle = -90,scale=0.32]{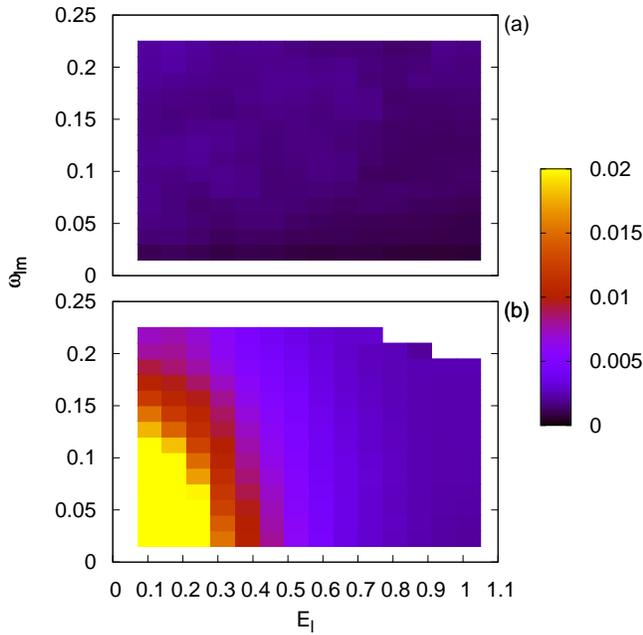} \\
\end{center}
\caption{(color online) Energy resolved current matrix element squared for(a) W=5 and (b) W=2 for a (16,20) graphene system subject to disorder.}
\label{fig:energyres}
\end{figure}

For weak disorder strength, we study the d.c. limit of average current matrix elements squared for different locations in the band. We consider a narrow band of states centred at different energy eigenvalues and calculate their contribution to $<|M_{lm}(\omega)|^{2}>$ in the limit $\omega \rightarrow 0$. It could be directly seen from the Fig.(\ref{fig:energyres}(a)) that  in the limit $\omega \rightarrow 0$ for $W$=5, contribution from all the energy states to conductivity is nearly equal. This implies that all the states are of the similar nature. However, in the same limit for $W$=2, the states close to the Fermi energy have a larger contribution and the contribution decreases as we move away from the band centre as seen in Fig. (\ref{fig:energyres}(b)). This implies that, close to band centre, the states are more extended (compared to the states away from the band centre). We shall return to the issue of localization in the next section.
\begin{figure}
\begin{center}
\includegraphics[angle = -90,scale=0.33]{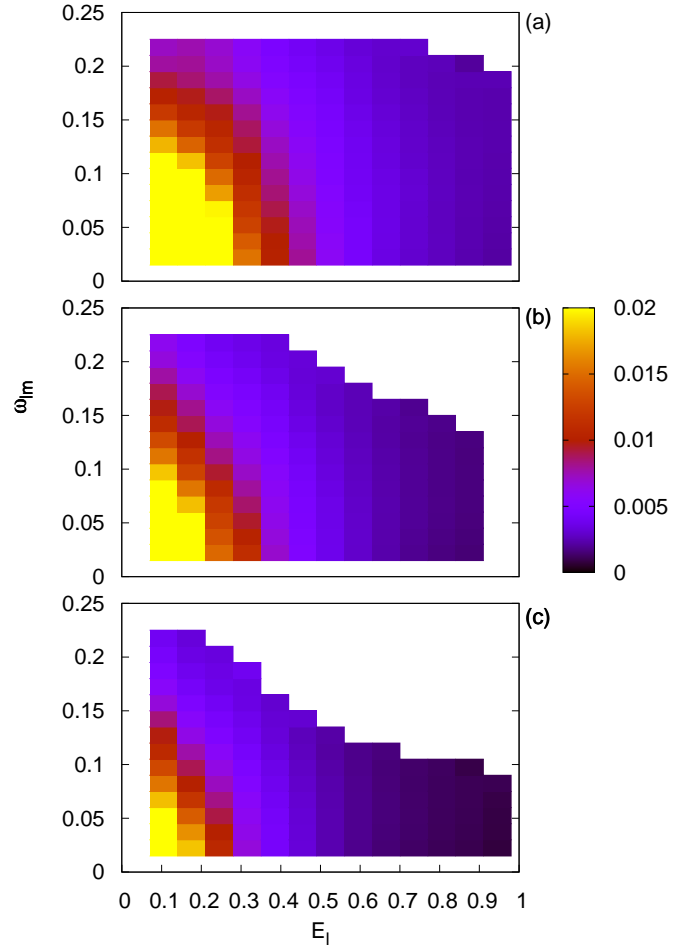}
\end{center}
\caption{(color online) Energy resolved current matrix element squared for W=2 for (a) (16,20), (b) (20,24) and (c) (24,30) graphene systems subject to disorder.}
\label{fig:energyres_W2}
\end{figure}
\begin{figure}
\begin{center}
\includegraphics[scale=0.73]{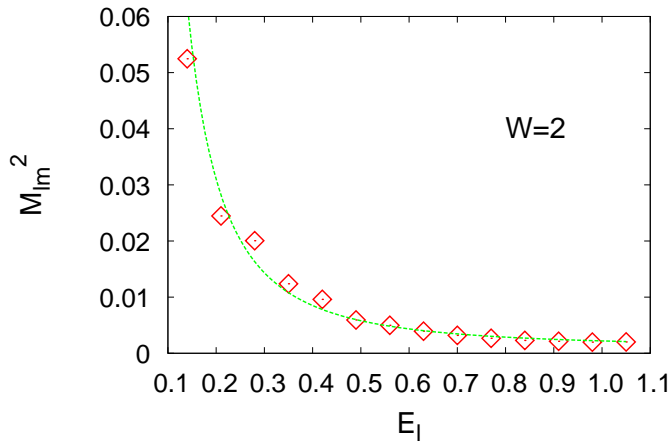}
\end{center}
\caption{(color online) The matrix element squared, $ {\vert M (\omega \rightarrow 0)\vert}^2 $ v/s $E_l$ for a (16,20) graphene lattice and W=2. The curve is a $a/E^{2}_{l} + b$ fit with a= 0.0010 and b = 0.002}
\label{fig:omega} 
\end{figure}

We may now ask : what is dependence of $\mid M_{lm}(\omega \rightarrow
0) \mid^{2}$ for $W$=2 on energy eigenvalues as we move along the band?
We polynomial fit the data and find (Fig.(\ref{fig:omega})) that the
behaviour is inversly proportional to the square of the energy
eigenvalue. The dependence is of the form $a/E_{l}^{2} + b$. For W =2 and system size (16,20), we find that $a = 0.0010$, and $b = 0.0021$.

\begin{figure}
\begin{center}
\includegraphics[angle = -90,scale = 0.32]{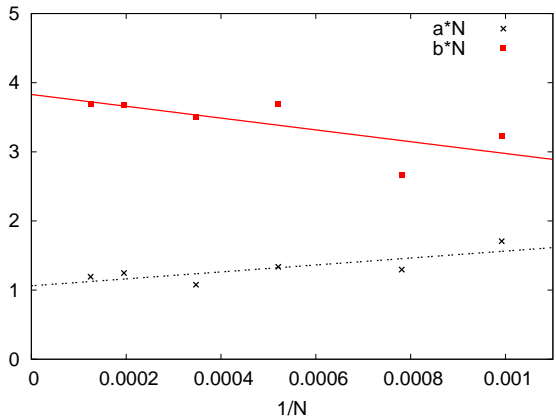}
\end{center}
\caption{(color online) Scaling of $aN$ and $bN$ with respect to system size.}
\label{fig:scaling}
\end{figure}

We study the scaling behaviour of the constants $aN$ and $bN$ with increasing system size. (Fig. \ref{fig:scaling}). We have plotted the behaviour of $aN$ and $bN$ versus $1/N$. It could be seen that there is not a significant change in the value of $aN$ and $bN$ as we increase the system size and the dependence of these quantities on $1/N$ is almost linear. As $N \rightarrow 0$, the values of $aN$ and $bN$ are 1.06157 and 3.8294 respectively.

The first term ($\sim 1/E_{l}^2$) is what is expected from the Boltzmann
theory. The other term, independent of $E_l$ is the non-Boltzmann
contribution.  A particle with the Hamiltonian described by
Eq.(\ref{tbham}), which supports both positive and negative energy
states(positive and negative bands), could be in a linear combination of
the positive and negative energy eigenstates. The Boltzmann term arises
out of transition between term in the same band. Off-diagonal terms in
the velocity operator correspond to the transition between the same
$\vec{k}$ but belonging to different bands and is responsible for the
non-Boltzmann contributions to the conductivity. The origin of this term
could also be understood from the fact that the velocity operator is
non-diagonal in the helicity basis. An analytical calculation for
Boltzmann conductivity with screened Coulomb scatterers and incorporating the
off-diagonal terms in the velocity operator was carried out by Trushin
\textit{et al.}\cite{Trushin}

\subsubsection{Conductivity and Mobility}
In order to study d.c. conductivity, we go back to Eq.(\ref{KG}) whereby d.c. conductivity could be written as the product of the current matrix elements squared in the limit $\omega \rightarrow 0$ and DOS squared. In order to do this, we polynomial fit the DOS squared and the average current matrix elements squared in the d.c. limit.

For $W$ =2, we have seen that DOS  $\sim \omega$. Carrier concentration,
$n$ is defined as $n = 3.8 \times 10^{14}$ $E_{F}^2 (cm ^{-2})$. Putting in the values of the parameters, $t$ = 2.8eV and $a$ = 1.42 $\AA$, together with the the bahavior of $\mid M_{lm}(\omega \rightarrow 0) \mid^{2}$ versus $E_l$ (obtained in the previous subsection) and $n$ versus $E_l$, we obtain conductivity as a function of the carrier concentration. Here the factor of two for spin multiplicity has already been taken into account. For large carrier concentration, we notice that as $E_l$ increases, the constant term ( "$b$" ) in the fit dominates the current matrix element contribution. This together with the DOS shows a linear dependence on the carrier concentration.  
\begin{figure}
\begin{center}
\includegraphics[angle = -90,scale = 0.33]{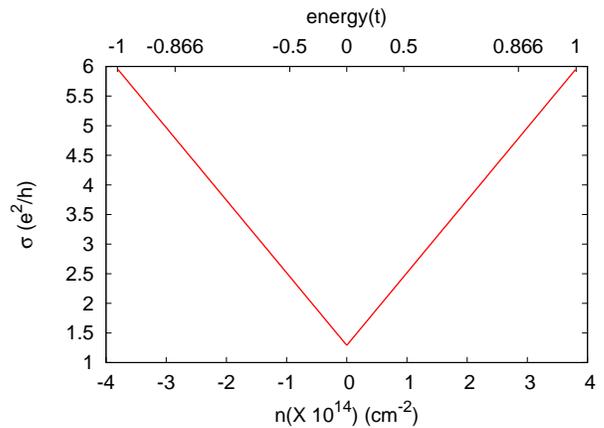}
\end{center}
\caption{(color online) Behavior of conductivity with carrier concentration for $W = 2$. The minimum value of conductivity at n = 0, $\sigma_{\rm min}$ is about $1.3e^{2}/h$.}
\label{fig:sigma_n}
\end{figure}

We also note that for n $\approx$ zero, the conductivity does not go to zero. It turns out that for small $W$, both the $|M_{lm}|^{2}$  and the DOS together conspire to give a non-zero d.c. conductivity ($\sigma_{\rm min}$), independent of $E_{F}$, close to the Dirac point. Comparision of different system size suggests that as the system size is increased the lesser number of eigensatates contribute to $\sigma_{\rm min}$ for very small doping concentration. However, the $\sim 1/E_{l}^2$ dependence is still retained. This translates to the fact that as we increase the system size, the $1/E_l^{2}$  shifts towards the Fermi energy as seen in Fig.(\ref{fig:energyres_W2}).

The value of this minimal conductivity, $\sigma_{\rm min}$  $\sim$ $4e^{2}/\pi h$ in confirmity with the previous results obtained with SCBA\cite{Fradkin1,Fradkin2,Castroneto2}.

We also compare this behaviour with respect to the system size and find that for larger system size, the number of such extended states decreases as shown in Fig. (\ref{fig:energyres_W2}) but still retaining its $\sim 1/E_{l}^2$ dependence. 

Thus, away from the zero doping concentration, one obtains a linear dependence of conductivity on the doping concentration in confirmity with the experimental findings. However, there is a quantitative difference between the two. The slope of the $\sigma$ v/s $n$ (Fig.(\ref{fig:sigma_n})) is very small compared to the experimental results.

The dependence if a.c. conductivity on $\omega$ can be obtained from the
energy resolved matrix elements squared. For a fixed energy, the
dependence of $|M_{lm}(\omega)|^2 \sim 1/\omega^2 $. This together with
the 

\begin{figure}
\begin{center}
\includegraphics[angle = -90,scale = 0.36]{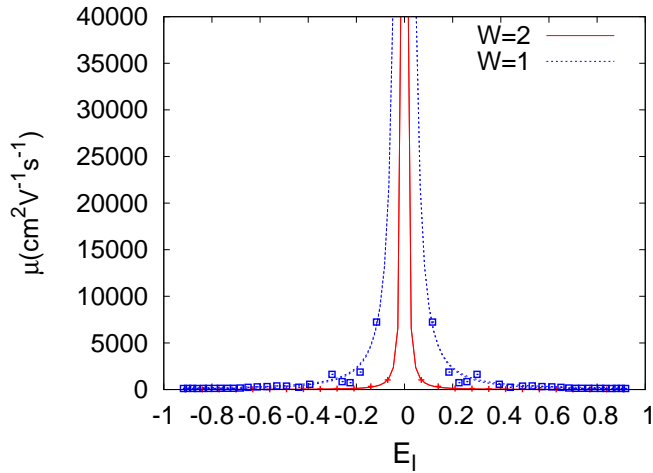}
\end{center}
\caption{(color online) Plot of mobility v/s gate voltage (carrier concentration) for a (20,24) graphene lattice for different disorder strengths $W$. The plot matches qualitatively with the experimental result.}
\label{fig:mobility}
\end{figure}
\emph{Mobility} is defined as: $\mu = \sigma/ne$. Since, $\sigma$ goes linearly as $n$, $\mu$ should nearly be a constant for energies away from the Dirac points. Since $n \propto E_{F}^2$, from Eq.(\ref{sigma}), we find that $\mu \propto |M_{lm}(\omega)|^{2}$. 

\begin{equation}
\mu = \frac{16 \pi^{2} N}{3 \sqrt{3}h \times 3.8 \times 10^{14}}(\frac{a}{\omega^{2}} + b)
\end{equation}
In Fig.(\ref{fig:mobility}), we plot $\mu$ versus gate voltage, $V_g$ where $n = 7.2 \times 10^{10} V_{g}$. We find that the mobility rises sharply close to the Dirac point. This sharp rise around the Dirac points is in a very good qualitative agreement with the experiments. However, the values of mobility is around 1000 times lower than the experimental results.

We would like to add that, in the adopted formalism, it is impossible to distinguish the effect of inter-valley scattering or to determine it's contribution towards conductivity. However, since we are dealing with very short range scatterers, it would be impossible to avoid large momentum interactions and hence inter-valley scattering.

\section{Localization}
\begin{figure}
\begin{center}
\includegraphics[scale = 0.55]{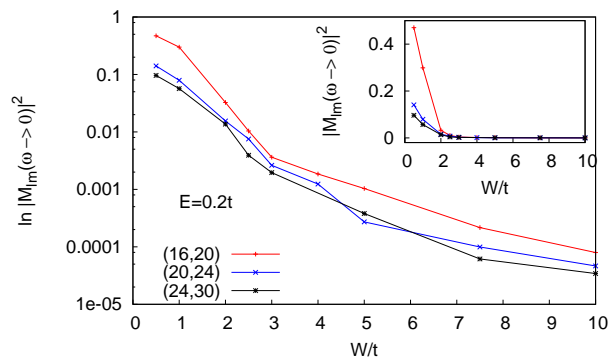}
\end{center}
\caption{(color online) Average current matrix elements square v/s W for different system sizes. The y-axis corressponds to the average dc value of for a band of ten states around the fixed energy 0.2$t$.}
\label{fig:M_W}
\end{figure}

In order to study localization, we looked at IPR versus energy eigenvaluei(not shown) and found that IPR is not sensitive enough to capture signatures of localization for small $W$. Therefore,we, again, turn towards the energy resolved studies of averaged current matrix elements squared, (Fig. \ref{fig:energyres}). As mentioned earlier, for a few states close to the band centre have a very large averaged current matrix element squared value. This implies that these state are extended. We notice that this is the case independent of the system size. Also, the experimental finding of a universal zero-bias minimum conductivity, $\sigma_{\rm min}$ suggests that the states close to the band centre should be extended, as we find which is in conformity with the findings of Amini \textit{et al.}\cite{Amini}. This, however, is in contradiction with the findings of Xiong \textit{et al.}\cite{Xiong}.

One might, also, question the $\mid M_{lm}(\omega) \mid ^{2}$ dependence on the energy eigenvalues, $E_l$ since it could very well be a finite-size effect. For low disorder strength, the localization length, $\xi$ varies exponentially with the inverse of disorder strength i.e. $\xi \sim exp({1/W})$ and in order to see the effects of localization, one needs to change the system size exponentially.We note that the $\sim 1/E^{2}_{l}$ behaviour is present at all system sizes starting from system containing nearly 300 lattice points((8,10)) to nearly 3000 lattice points((24,30)). 

Also, that for $W$=3, the $|M_{lm}(\omega)|^2$ is independent of $E_l$ (not shown) suggesting a disorder induced crossover from localized to delocalized phase for $2 \le W \le 3$. We check this by studying the averaged current matrix element squared over a range of $W$ for different system sizes, at a fixed energy. We have fixed the energy at $0.2t$ and have considered ten states around this energy. We study the contribution of these ten states towards the $|M_{lm}(\omega)|^2$. The results are shown in Fig.(\ref{fig:M_W}). The inset of Fig.(\ref{fig:M_W}) shows the plot between averaged current matrix element squared over a range of $W$ for three different system sizes (16,20),(20,24) and (24,30). The cross-over is highlighted in the main panel which shows the logarithm of averaged current matrix elements squared versus $W$. We indeed find a crossover around $W$=3. However,this crossover is not between localized and delocalized phases but from localized to a very weakly localized behaviour. This is in agreement with the scaling theory of localization which predicts that all states in 2D are localized.

\section{comparision with square lattice}
\begin{figure}
\begin{center}
\includegraphics[angle = 0,scale=0.6]{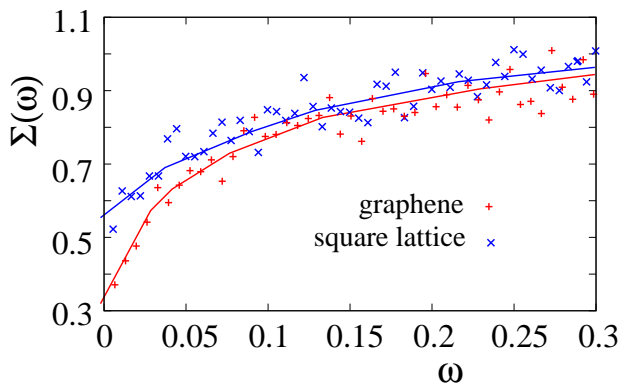}
\end{center}
\caption{(color online) Comparision of $|M|^{2}$ values between a square lattice of size (36,36)(=1296 lattice points) and graphene of size (16,20) (=1280 lattice points) for W=5. Averaging has been done over 50 configurations of disorder.}
\label{fig:comparsq} 
\end{figure}
We, now, compare the normalized conductivity, $\Sigma(\omega)$ between square lattice and graphene for $W$ = 5. This comparision has been shown in Fig.(\ref{fig:comparsq}). We observe that the conductivity matches well at large frequencies for both the lattices. However, at low frequencies, the grahene has a lower conductivity suggesting that the states are more localized. This could mean that the states in graphene are more susceptible to localized when subject to disorder, as also suggested by Xiong \textit{et al.}\cite{Xiong}
 
\begin{figure}
\begin{center}
\includegraphics[angle = -90,scale=0.18]{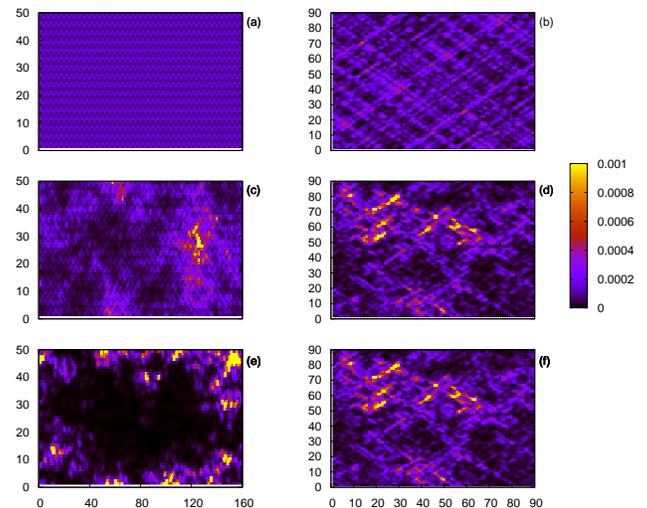}
\end{center}
\caption{(color online) Intensity plot of energy eigenstate next to the band centre for pure graphene sheet ((a),(c) and (e)) and square lattice ((b),(d) and (f))for different values of disorder strength viz. disorder-free (pure) ((a) and (b)), W=2 ((c) and (d))and W=5 ((e) and (f)). The system sizes considered  are comparable for both the systems, consisting of 8000 ((40,50)) and 8100 ((90,90)) lattice points.}
\label{fig:comparsq2}
\end{figure}

We also compare the intensity plots for graphene and square lattice for different values of disorder strengths and fixed system size. The results are shown in FIG(\ref{fig:comparsq2}). We find that there is no signature fo puddle formation for square lattice. The reason for the formation of puddle for low carrier concentration in graphene is yet to be understood.

\section{Conclusion}

In conclusion, we have studied the finite-frequency electrical
conductivity in graphene under diagonal (on-site) disorder of various
strength and in the presence of inter-valley scattering using
Kubo-greenwood foumula. For moderate disorder strength, we find that for
differnt system sizes, fixed $W$, scaling theory is at work. We made
contact with the weak localization result of Abrahams \textit{et
al.}\cite{Abrahams} and found a very good agreement which means that for
$W=5$, logarithmic scaling is obeyed. We compare normalized conductivity of
graphene with that of a square lattice and find that graphene is more
susceptible to localization when subject to disorder. We have
established that, for low disorder strength, $W=2$, the states away from
the band centre are more localized comared to the ones close to the band
centre. For low disorder strength, we have calculated the conductivity
for low-disordered graphene and have found the results are in
disagreement with that of Boltzmann conductivity.
Also, the conductivity and mobility are in qualitative
agreement with experiments. Also, for weak disorder strength, it is the
competition between $|M(\omega)|^{2}$ and DOS which gives rise to a
universal conductivity minimum. The linear rise of conductivity with
carrier concentration is due to a term unaccounted for in the Boltzmann
expression for conductivity. We have also established that, for low
disorder strength, $W=2$, the states close to the band centre are
extended and that there exists a crossover at $W$ $\sim$ 3. Comparative
study of intensity plots for states close to the band centre for
graphene with square lattice shows clear signatures of puddle formation in graphene.
Although we have studied a simple disorder model in graphene,some of the feature studied could be generic.

\end{document}